\documentclass[12pt]{article} 
\usepackage{amsfonts, amssymb, amsmath}
\usepackage{graphicx,epstopdf}

\def\be{\begin{eqnarray}}
\def\ee{\end{eqnarray}}
\def\bee{\begin{eqnarray*}}
\def\eee{\end{eqnarray*}}
 \def\pmx{\begin{pmatrix}}
 \def\emx{\end{pmatrix}}
 \def\bsq{\begin{subequations}}
\def\esq{\end{subequations}}

\newtheorem{thm}{Theorem}
\newtheorem{cor}[thm]{Corollary}
\newtheorem{lemma}[thm]{Lemma}

\def\pf{\noindent {\bf Proof: }}
\newcommand{\norm}[1]{ \| #1  \|}

        \def\tr{\hbox{Tr} \, }
     \def\trp{\hbox{Tr} }
     
     \def\half{{\textstyle \frac{1}{2}}}
     \def\nn{\nonumber}
     \def\rank{{\rm rank}}
          
\def\cB{{\cal B}}

\def\cE{{\cal E}}

\def\cP{{\cal P}}
\def\cS{{\cal S}}

\def\cH{{\cal H}}
\def\cD{{\cal D}}
\def\cV{{\cal V}}
\def\pp{ \! + \! }
\def\mm{ \! - \! }
\def\e{\epsilon}

\def\ds{\displaystyle}
 
\def\bra{\langle}
\def\ket{\rangle}
\def\kb{ \ket \bra }

\def\rt2{ \frac{1}{\sqrt{2}} }

\def\raw{\rightarrow}

\def\wh{\widehat}
           \def\wtd{\widetilde}
\newcommand{\proj}[1]{ | #1 \kb  #1|}
\newcommand{\ovb}[1]{\overline{ #1 }}
   \def\qed{{\bf QED}}

\def\ot{\otimes}

\title{Bipartite states of low rank are  \\ almost surely entangled}

\author{Mary Beth Ruskai \thanks{Partially supported by  National Science Foundation 
 under Grant DMS-0604900.}
 \\{\small  Department of Mathematics, Tufts University, Medford,
MA 02155, USA} \\ {\small \tt marybeth.ruskai@tufts.edu} 
\and Elisabeth M. Werner\thanks{Partially supported by an NSF grant, a FRG-NSF grant and  a BSF grant}\\ {\small Department of Mathematics, Case Western Reserve University} \\
{\small Cleveland, Ohio 44106, U. S. A. } \\
{\small Universit\'{e} de Lille 1, UFR de Math\'{e}matique, 59655 Villeneuve d'Ascq, France}\\
{\small \tt elisabeth.werner@case.edu}}

\begin{document}

\maketitle

\begin{abstract}
We show that a bipartite state on  a tensor product of two matrix 
algebras is almost surely entangled if its rank is not greater than
that of one of its reduced density matrices.   
\end{abstract}

\section{Introduction}

\subsection{Background}

Recently, Arveson \cite{Arv} considered the question of when a bipartite mixed state 
of rank $r$ is almost surely entangled, and showed that this holds
when $r \leq d/2$ where $d$ is the dimension of the smaller space.    
In this note we show that this result holds if $r \leq d$, with $d$ now
the dimension of the larger space.

We will use results from \cite{HSR} on entanglement breaking channels and
 exploit the well-known isomorphism between bipartite states and 
completely positive  (CP) maps.\footnote{This isomorphism is usually attributed
to Jamiolkowski \cite{J} or to Choi \cite{C}, who used it to characterize 
the complete positive maps on finite dimensional algebras.   However, it seems to
have been known to operator algebraists earlier and appeared implicitly
in Arveson's proof of Lemma~1.2.6 in \cite{A69}  }.   
We will first consider states associated with completely positive  trace-preserving
(CPT maps) and then find that
extension to arbitrary bipartite states is quite straightforward.

If the rank of a  bipartite state $\gamma_{AB}$ is
strictly smaller than that of either of its reduced density matrices, then the
state must be entangled.   This is an immediate consequence of 
well-known results on entanglement, and  seems to have first appeared explicitly 
in \cite{HSTT}.    We include a proof in Appendix~\ref{sect:A} for completeness.   This allows
us to restrict attention to the case in which the ranks of the reduced density matrices
are equal, with one of full rank.

Although it seems natural to expect that this result is optimal, recent results
of Walgate and Scott \cite{WS}  suggest otherwise.   Let the Hilbert spaces
 $ \cH_A$ and $\cH_B$ have dimensions $d_A$ and $d_B$ respectively.
It follows from a result proved independently by  Wallach \cite{W} and by Parthasarathy  \cite{P}  
for multi-partite entanglement that when $s > (d_A - 1)(d_B -1) $ any subspace of 
$\cH_{AB} = \cH_{A} \ot \cH_B $  with dimension $s$ contains some
product states, and that this bound is best possible, i.e., if $s \leq  (d_A - 1)(d_B -1) $
then there is some subspace of dimension $s$ with no product states.
  
      Walgate and Scott extended this by proving  \cite[Corollary~3.5]{WS} 
that if a subspace of  $ \cH_{A} \ot \cH_B $ has dimension $s \leq (d_A - 1)(d_B -1) $
then, almost surely, it contains no product states.    For a bipartite state
$\gamma_{AB}$ with rank $ r \leq (d_A - 1)(d_B -1) $, it follows that range of
 $\gamma_{AB}$ almost surely  contains no product states,  which implies that a
bipartite state $\gamma_{AB}$ with rank $r \leq (d_A - 1)(d_B -1) $ is almost
surely entangled.      Alternatively, one could apply \cite[Theorem~3.4]{WS} 
directly to $\ker(\gamma_{AB}) $ to reach the same conclusion.
 
  When $d_A >  d_B \geq 2$, this  result is
 stronger than ours, but for a pair of qubits, $d_A = d_B = 2$  our result
  is stronger.    Moreover, it is easy to extend our qubit results to the general
  case of bipartite states with rank $r = d_A \geq  d_B \geq 2$,
  providing a proof quite different from that in \cite{WS}.
 Although our measure is constructed differently from that used in \cite{Arv},
our approach is similar in the sense that we show that in a natural parameterization
of the set of density matrices, the separable ones lie in a space of smaller
 dimension.   

In the next half of this section, we review relevant terminology,   
and describe the notation and conventions we will use.
 Qubit channels and states are considered in Section~\ref{sect:qubit}, and
 the general case in Section~\ref{sect3}.    We conclude with some remarks  
about other approaches, and the question of the largest
rank for which the separable states have measure zero.


\subsection{Basics and notation}

In this paper, we consider maps   $\Phi : \cB(\cH_A) \mapsto  \cB(\cH_B)$
and identify them with bipartite states or, equivalently, density matrices in 
$ \cB(\cH_A) \ot  \cB(\cH_B)$ via the Choi-Jamiolkowski isomorphism
as described below.
Our primary interest is the situation in which $\cH_A =  {\bf C}_{d_A}$,
in which case we can identify  $\cB({\bf C}_{d })$ with $M_{d }$,
the space of $d \times d $ matrices. 
However, we will also have occasion to consider either Hilbert space
$\cH$ as a proper subspace of ${\bf C}_d$ for some $d$.
 
   We will identify a state with
a density matrix, i.e., a positive semi-definite operator $\rho$ with
$\tr \rho = 1$, {\em in} $ \cB(\cH)$.    To an operator algebraist
this corresponds to the positive linear functional {\em on} the algebra 
$\cB(\cH)$ which takes $A \mapsto \tr \rho \, A$.    In the physics and
quantum information literature, a density matrix  (or, more properly,
a density operator) is often referred to as a (mixed) state {\em on} $\cH$ 
(because the density operator acts on  $\cH$. )

When $\cH_A =  {\bf C}_{d_A}$ and $\cH_B =  {\bf C}_{d_B}$, we  write
 $\Phi: M_{d_A} \mapsto M_{d_B}$. 
In this case, 
   let  $\{ e_j \} $ and $\{ f_m \} $ denote orthonormal bases
for $ {\bf C}_{d_A}$ and $ {\bf C}_{d_B}$  respectively.  The isomorphism 
between states and matrices arises from the fact that
\be
        \tr   |f_m \kb f_n | \, \Phi(  |e_j \kb e_k|)
\ee
can be interpreted as either 

 (i)  the matrix representative of the linear map $\Phi: M_{d_A} \mapsto M_{d_B}$
 in the bases  $  |f_m \kb f_n |  $ and  $ |e_j \kb e_k| $ for $M_{d_B}$ and   $M_{d_A}$ 
 respectively, or,
 
 (ii)  the density matrix $\gamma_{AB} $ of a state  on $  {\bf C}_{d_A} \ot    {\bf C}_{d_B}$
  with elements  $[\gamma_{AB}]_{jm,kn}$  in the product basis $|e_j \ot f_m \ket$.
 
 Conversely, any state on $  {\bf C}_{d_A} \ot    {\bf C}_{d_B}$ defines a CP map.  
 We describe this well-known fact in detail in order to establish some conventions
 for interpretations of $\gamma_A $ and $\gamma_B$.
  Observe that (ii) is equivalent to writing $\gamma_{AB} $ as a block
 matrix of the form
 \be
   \gamma_{AB} =  \tfrac{1}{d_A} \sum_{jk} {|e_j \kb e_k|} \ot   P_{jk} =
       \tfrac{1}{d_A}   \sum_{jk} {|e_j \kb e_k|} \ot  \Phi(|e_j \kb e_k|)
 \ee
 with the block $P_{jk} = \Phi(|e_j \kb e_k|)$ the matrix in $M_{d_B}$ given 
 by the image $ \Phi(|e_j \kb e_k|)$.   One can write an arbitrary matrix in
 $M_{d_A} \ot M_{d_B}$ in the block form  $ \sum_{jk}  |e_j \kb e_k| \,   P_{jk} $
 and then define  $ \Phi(|e_j \kb e_k|) = P_{jk}$ and extend by linearity or, equivalently, 
 \be
        \Phi(A) = \sum_{jk} a_{jk} P_{jk}
 \ee
 when $A = \sum_{jk} a_{jk} |e_j \kb e_k| $.  
 
Observe that 
\bsq \label{rdm}  \be   \label{rdma}
     \gamma_B & = &  \tfrac{1}{d_A} \trp_A  \gamma_{AB} =  \tfrac{1}{d_A} \sum_k \Phi( \proj{e_k}) =  
        \tfrac{1}{d_A}  \Phi(I_A)\\
     \label{rdmb}
    \gamma_A  & = &  \tfrac{1}{d_A}  \trp_B  \gamma_{AB}  = 
             \tfrac{1}{d_A}\sum_{jk}   |e_j \kb e_k|  \, \tr \Phi(|e_j \kb e_k|)  
\ee  \esq
and that this implies the following:

a)   $\Phi$ is unital, i.e.,  $\Phi(I_A) = I_B$, if and only if  $\gamma_B =  \tfrac{1}{d_A} I_B$, and

b)  $\Phi$ is trace-preserving (TP), i.e., $\trp_B \, \Phi(X) = \trp_A \, X ~~\forall ~ X \in \cB(\cH_A)$,
 if and only if  $\gamma_A =  \tfrac{1}{d_A} I_A$.
 
 
When $M_d$ or $\cB(\cH)$ is equipped with the Hilbert-Schmidt inner product,
one can define the adjoint, or dual, of a map $\Phi$.   We denote this by
$\wh{\Phi}$ and observe that this is equivalent to
\be
      \tr B^\dag \Phi(A) = \tr [\wh{\Phi}(B)]^\dag A.
\ee
 A matrix $\Phi$ is TP if and only if its adjoint $\wh{\Phi}$ is unital.

It is a consequence of Theorem~5 in \cite{C} that the extreme points\footnote{Choi's 
condition for true extreme points is implicit in Theorem~1.4.6 of \cite{A69}.}   of the
convex set of CP maps for which $\gamma_A =  \wh{\Phi}(I_B) = \rho$ 
have a state representative (often called the Choi matrix)  with rank $\leq$ rank $\rho$.   
We prefer to consider CPT maps and 
regard the density matrices with rank $\leq d_A$ as an
 extension of the set of extreme points.  As shown in Appendix~\ref{app:ext},
   this corresponds to
 the closure of the set of of extreme points.  We let  $\cD_C$ denote the set
 of density matrices in  $ \cB(\cH_C)$  or $M_{d_C }$ and $\cD_C(r)$ to denote
 the subset of rank $r$.   We also
 define the  following subsets   of $\cD_{AB}(r)$.
\bsq   \label{Ddef} \be
   \cP_A(\rho;r,s)  & \equiv & 
     \{  \gamma_{AB} \in {\cal D}_{AB} :  \rank \, \gamma_{AB} = r ,  ~ 
    \rank \, \gamma_{A} = s ~\hbox{and} ~ \gamma_A = \rho  \} \\
    \cP_A(r,s)   & \equiv &    
      \{  \gamma_{AB} \in {\cal D}_{AB} :  \rank \, \gamma_{AB} = r ~\hbox{and} ~ 
          \rank \, \gamma_{A} = s \} 
 \ee \esq
  Although the sets in \eqref{Ddef} above  are
 subsets of ${\cal D}_{AB} \subset   \cB(\cH_A) \ot \cB(\cH_B) \simeq M_{d_A} \ot M_{d_B} $ 
 we use the subscript $A$
 to emphasize that we impose conditions only on the marginal $\gamma_A$.
 When   rank $\rho_1 = $ rank $\rho_2 = d_A$, the map
 \be \label{rhoiso}
 \gamma_{AB}  \mapsto  ( \rho_2^{1/2}  \rho_1^{-1/2} \ot I_B )  \gamma_{AB} \, 
     ( \rho_1^{-1/2} \rho_2^{1/2}  \ot I_B  )  
  \ee
 gives an isomorphism from $ \cP_A(\rho_1;r,d_A) $ to $\cP_A(\rho_2;r,d_A) $
 and each of these  is isomorphic to   
$\cP_A(\tfrac{1}{d_A} I_A ; d_A,d_A)$ which is isomorphic to the set of 
  CPT maps $\Phi$ whose Choi matrix has rank  $d_A$.   We will let $ \cS_A(\rho;r,s) $, etc.  denote the
 corresponding  subsets of separable  state in \eqref{Ddef}.
 
 It will be useful to introduce the notation $\Upsilon_T$ for the map that takes
 a density matrix $\rho  \mapsto  T^\dag \rho T$.   
 
 \pagebreak

 \section{Maps with qubit inputs}   \label{sect:qubit}
 
 \subsection{Canonical form and parameterization}  \label{sect:qq}
 
 Now consider the case of CPT maps on qubits for which $\cH_A = \cH_B = {\bf C}_2$.
As observed in \cite{KR1}, these maps can be written using the Bloch sphere
representation in the  form 
\be   \label{KRrep1}
    \Phi\big( w_0 I + \sum_k w_k \sigma_k \big) = w_0 I + \sum_k (t_k w_0 + \lambda_k w_k ) \sigma_k
\ee
 where $\sigma_k$ denote the three Pauli matrices.   Necessary and sufficient
 conditions on $t_k, \lambda_k$ which ensure that $\Phi$ is CP are given in \cite{RSW}.
The form \eqref{KRrep1} is equivalent  to representing $\Phi$ by a matrix $T$
with elements  $t_{jk} = \half \tr \sigma_j \Phi(\sigma_k)$ so that, with
subscripts $j,k = 0,1,2,3$ and the convention $I_2 = \sigma_0$ 
\be   \label{KRrep2}
       T = \pmx 1 & 0 & 0 & 0\\ t_1 & \lambda_1  & 0 & 0 \\
                t_2 & 0 &  \lambda_2 & 0 \\  t_2  & 0 & 0  & \lambda_3 \emx.
\ee
As shown in \cite[Appendix B]{KR1} an arbitrary 
unital  map on qubits can be reduced to this
form by applying a variant of the singular value decomposition to the $3 \times 3$
submatrix  with $j,k \in \{ 1,2,3 \}$ using only real orthogonal rotations.    Given 
the isomorphism between rotations and $2 \times 2$ unitary matrices, this
corresponds to making a change of basis on the input and output spaces
${\cal H}_A =    {\bf C}_{d_A} = {\bf C}_2$ and $  {\cal H}_B = {\bf C}_{d_B} =  {\bf C}_2 $
respectively.    Thus, for an arbitrary unital CP map $\Phi$ one can find unitary
$U,V$ such that  $\Upsilon_{V^\dag} \circ \Phi \circ \Upsilon_U$ has the form
\eqref{KRrep1} or, equivalently, a matrix representative of the form \eqref{KRrep2}.

    It was shown in \cite{RSW} that the maps with Choi rank $\leq 2$
are precisely those for which the form \eqref{KRrep2} becomes
\be   \label{RSWrep}
   T_{u,v} = \pmx 1 & 0 & 0 & 0 \\ 0 & \cos u  & 0 & 0 \\
                0 & 0 &  \cos v & 0 \\   \sin u \sin v  & 0 & 0  & \cos u \cos v \emx
\ee
with\footnote{The interval for  $u$ is shifted from that in \cite{RSW}.   However,
the interval $[0,\pi]$  for $v$ was incorrectly stated as $[0,\pi)$ in \cite{RSW}.}
 $u, v$ in  $(-\pi,\pi] \times [0,\pi] $.   Moreover, as shown in \cite{EBQ}, the 
entanglement breaking (EB) maps    are precisely the channels which have either
$\cos u = 0$ or $\cos v = 0$.

\begin{figure}[h]   \label{fig1}
 
 \begin{center}
  \includegraphics*[width=6cm,height=6cm,keepaspectratio=true]{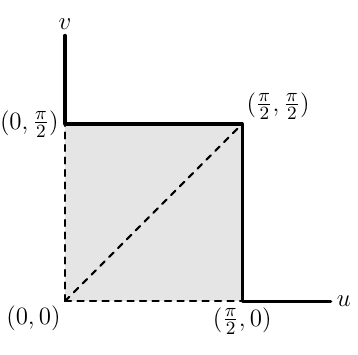}   
  \end{center}
  
 \caption{The rectangle $\ovb{\Delta}$ corresponds to the shaded region.   
 The dashed lines are not in $\Delta$.  
 The lines   $ u =  \frac{\pi}{2}$ and $ v = \frac{\pi}{2}$ correspond to the EB channels. }
 
 \end{figure}

It follows from \eqref{RSWrep} that every element of $\cP_A(\half I; 2,2) $
can be represented by a triple $\big( (u,v) , U, V \big)$ consisting of a point in ${\bf R}_2$,
and two unitary matrices $U, V$.   However, some care must be taken so that
each element of $\cP_A(\half I; 2,2) $ is counted exactly once.   It suffices to
restrict $(u,v)$ to the rectangle
\be  \label{tribar}
   \ovb{ \Delta} =  \big[0,  \frac{\pi}{2} \big] \times \big[0,  \frac{\pi}{2} \big] 
    \ee    
 Suitable  rotations will give all allowed negative
values of the non-zero elements in \eqref{RSWrep}, as well as even
permutations of $t_k$ and $\lambda_k$.    Problems with 
overcounting occur only on the lines $u = 0, v = 0, u = v$.  To deal with this we define
 \be  \label{tri}
   \Delta  = \{ (u,v) :  0 < u \leq \frac{\pi}{2}, 0 < v \leq \frac{\pi}{2}, u \neq v \} .
    \ee      
    (The line segments  on the boundary with  $u = \frac{\pi}{2} $ and $ v = \frac{\pi}{2} $ 
 are   included in $\Delta$ as shown in Figure~\ref{fig1}.)   
 
 Because different pairs of matrices $U, V$ may
 give the same channel on the lines not included in \eqref{tri}, 
 we define equivalence classes as follows.
  Let $ {\cal R}_t $  (with $t = x,y,z$) denote the subset of $SU(2)$
 corresponding to the rotations around the indicated axis.   
 We write  $(U,V) \simeq (U^\prime, V^\prime)$  if
there is an $R_t \in  {\cal R}_t $ such that $U^\prime = R_tU$ and
$V^\prime = R_t V$ or, equivalently 
$U^\prime U^\dag =  V^\prime V^\dag \in {\cal R}_t $, and
denote the quotient space  $(SU(2) \times SU(2))/{\cal R}_t $.
With this notation, we now make some observations
\begin{itemize}

\item[a)]  The subset of EB channels consist of those channels for which either
  $ u =  \frac{\pi}{2}$ or $ v = \frac{\pi}{2}$;

\item[b)]   The line $u = v$ corresponds to the  amplitude damping channels
   (It is well-known that only the case $u = v = \frac{\pi}{2}$ is EB; this
    is a completely noisy channel mapping to a fixed pure state.)
   From \eqref{RSWrep} one sees that these channels are invariant under rotations 
   about the $z$-axis,  
   and the  set of amplitude damping channels in $\cP_A(\half I; 2,2) $ is isomorphic to
  $ (u,u) \times (SU(2) \times SU(2))/{\cal R}_z $.
   
 \item[c)]  The line segments with $u = 0$ and $v = 0$ correspond to
 phase-damping channels.    
    From \eqref{RSWrep} one sees that these channels are invariant under rotations 
   about the $x$ and $y$-axes respectively.   Thus, the
    set of phase damping channels in $\cP_A(\half I; 2,2) $ is isomorphic to
 \bee
\lefteqn{  \{ (0,v) : v \in (0,\tfrac{\pi}{2},] \} \times   (SU(2) \times SU(2))/{\cal R}_x } \qquad \\
  &  \bigcup  & 
    \{ (u,0) : u \in (0,\tfrac{\pi}{2}] \} \times   (SU(2) \times SU(2))/{\cal R}_y .
  \eee
    
 \item[d)]   The point  $u = v = 0$ gives the identity channel, for which rank $\gamma_{AB} = 1$.
 
 \end{itemize}

  Thus $\cP_A(\half I; 2,2) $ is isomorphic to
  \be   \label{cpt2}
      \Delta \times SU(2) \times SU(2)   \,  \bigcup  \, 
     \{ (u,u) \}_{u \in (0,\frac{\pi}{2}]} \times  (SU(2) \times SU(2))/{\cal R}_z    \quad \nn \\   
     \bigcup \,  \{ (u,0) \}_{u \in (0,\frac{\pi}{2}]} \times  (SU(2) \times SU(2))/{\cal R}_y    \\ \nn 
        \bigcup \,  \{ (0,v) \}_{v \in (0,\frac{\pi}{2}]} \times  (SU(2) \times SU(2))/{\cal R}_x     
        \ee
and, $\cS_A(\half I; 2,2) $, the subset of EB channels in $\cP_A(\half I; 2,2) $, is isomorphic to
    \be  \label{eb2}
    \lefteqn{    \{ (u,\tfrac{\pi}{2}) : u \in (0,\tfrac{\pi}{2}) \} \times SU(2) \times SU(2) } \nn  \\
    & ~ &  \,   \bigcup  \,  \{ ( \tfrac{\pi}{2}, v) : v \in (0,\tfrac{\pi}{2}) \} \times SU(2) \times SU(2)    
        \bigcup \, (0,  \frac{\pi}{2})  \times  (SU(2) \times SU(2))/{\cal R}_x   \quad \nn \\
     & ~ &     \,   \bigcup \,   ( \frac{\pi}{2}, 0)  \times  (SU(2) \times SU(2))/{\cal R}_y  
           \,   \bigcup \,       (\frac{\pi}{2} , \frac{\pi}{2}) \times  (SU(2) \times SU(2))/{\cal R}_z  .
             \ee
 

\subsection{Construction of a measure}   \label{sect:meas2}

Let $m_2$ be the normalized Lebegue measure on $\ovb{\Delta}$ and $\nu_2$ the normalized 
Haar measure on $SU(2)$.   Then the product measure
$\wtd{\mu} \equiv m_2 \times \nu_2 \times \nu_2$ defines a probability measure on 
$\Omega_2= \ovb{\Delta} \times SU(2) \times SU(2)$.
Although every point in $\Omega_2$ corresponds to an element in   $\cP_A(\half I; 2,2) $,
it can happen, as described above, that more than one point corresponds to the same
  CPT map $\Phi$.   Therefore, 
to define a measure on $\cP_A(\half I; 2,2) $
we  use the map  $g: \Omega_2 \rightarrow\cP_A(\half I; 2,2) $
which takes
\be
    \big( (u,v), U, V \Big) \mapsto  \Upsilon_V \circ \Phi_{u,v} \circ \Upsilon_{U^\dag}
\ee
where  $ \Phi_{u,v} $ denotes the CPT map whose Choi matrix is given by \eqref{RSWrep}.
The map $g$  is surjective which allows us to define a measure $\mu$ on all sets
$X \subset  {\cal P}_A(\half I; 2,2) $ for which $ g^{-1}(X)$
is measurable by
 \be   \label{mudef}
        \mu(X) = \wtd{\mu}\left(g^{-1}(X)\right).
\ee
  Since $g$ is surjective, 
$g^{-1}(\cP_A(\half I; 2,2) ) = \Omega_2$
which implies that $\cP_A(\half I; 2,2)$ is measurable and $\mu(\cP_A(\half I; 2,2) =1$.
Thus, $\mu$ is a probability measure on $ \cP_A(\half I; 2,2) $

Moreover,  the entanglement breaking channels satisfy
\be   \label{02}
\lefteqn{ \mu(   \cS_A(\half I; 2,2) )}  \nn  \\  \nn
&  = & \wtd{\mu} \big(
\{ (u,\tfrac{\pi}{2}) : u \in [0,\tfrac{\pi}{2}]  \} \times SU(2) \times SU(2)  \cup
    \{ ( \tfrac{\pi}{2}, v) : v \in [0,\tfrac{\pi}{2} \} \times SU(2) \times SU(2)      \big) \\
    & = &  0 \cdot 1 \cdot 1 + 0 \cdot 1 \cdot 1  = 0
     \ee

Thus, we have proved the following
\begin{thm}   \label{lemm22}
A CPT map $\Phi: M_2 \mapsto M_2$ of Choi-rank 2 is almost surely not EB, or, equivalently,
a state $\gamma_{AB}$ on ${\bf C}_2 \ot {\bf C}_2$ which has rank 2 and 
$\gamma_A = \half I$ is almost surely entangled.
\end{thm}

Since, the unitary conjugations have Choi matrices of rank  1, and
correspond to the set $(0,0) \times SU(2)$ which has
measure zero, we have also proved the following result, which we state
for completeness.
\begin{thm}   \label{lemm22}
A CPT map $\Phi: M_2 \mapsto M_2$ of Choi-rank $\leq 2$ is almost surely not EB, or, equivalently,
a state $\gamma_{AB}$ on ${\bf C}_2 \ot {\bf C}_2$ which has rank $\leq 2$ and 
$\gamma_A = \half I$ is almost surely entangled.
\end{thm}


\subsection{Removing the TP restriction} \label{sect:TP2}

We would like to extend the results of the previous section to
\begin{thm}  \label{thm22}
If a state $\gamma_{AB}$ on ${\bf C}_2 \ot {\bf C}_2$ has rank 2 and
$\gamma_A $ also has rank 2, then $\gamma_{AB}$ is almost surely entangled.
\end{thm}
\pf As observed after
\eqref{rhoiso}, $ \cP_A(\rho_1;r,d_A) \simeq\cP_A(\rho_2;r,d_A) $;
Indeed, the CP{ maps corresponding to states in    $\cD_A(\rho; 2,2) $ have the form
$\Phi \circ \Upsilon_ {\sqrt{ 2 \rho}} $ with $\Phi$ CPT, although it might seem
more natural to consider the dual  $\Upsilon_ {\sqrt{ 2 \rho}} \circ \wh{\Phi}$ which
takes $I \mapsto d_A \, \rho$.   
Next, observe that any density matrix $\rho  \in M_{d_A}$ of rank 2,  
can be written as $ U \left( \begin{smallmatrix} x & 0 \\ 0 & 1-x \end{smallmatrix} \right) U^\dag$ 
with $x \in (0,\half)$ and $U \in SU(2)$;
the  case $x = \half$ gives $\half I$ independent of $U$.  Thus
the set of density matrices $\rho  \in M_{d_A}$ of rank 2 is 
isomorphic\footnote{Here we use the fact that $\sigma_x \rho \sigma_x$
exchanges the eigenvalues.   This is quite different from the situation in
\eqref{tri} where we could not assume $u < v$ because the permutation in $S_3$
which exchanges $1 \leftrightarrow 2$ can not be implemented with a rotation.}
 to  
\be   \label{rho2}
     \half I  ~ \cup  ~  (0, \half)  \times SU(2)
\ee
and the set of bipartite density matrices  $ \cP_A(2,2)$ (for which rank~$\gamma_{AB} = $ 
rank $\gamma_A = 2$)  is isomorphic to
\be   \label{gen22}
  \cP_A(\half I; 2,2)  \bigcup    
        \cP_A(\half I; 2,2) \times (0,\half) \times SU(2).
\ee
To define a measure on this set, let  $m_1$ denote
 normalized Lebesque measure on 
  $(0,\half)$  and let $\lambda_{2,t}$ be defined using product measure so that
  \be   \label{m:gen2}
      \lambda_{2,t}(X) =  \begin{cases} t  \, ( \mu \times m_1 \times  \nu_2)(X) & ~~X \in 
          \cP_A(\half I; 2,2) \times (0,\half) \times SU(2) \\
         (1 \mm t)  \,  \mu(X) & ~~X \in  \cP_A(\half I; 2,2)    \end{cases}
  \ee
  where we can pick any $t \in (0,1]$ and $\mu $ is the measure defined in Section~\ref{sect:meas2}.
Then the subset of EB channels  $\cS_A(2,2)$ has measure
\be
   \lambda_{2,t} \big(  \cS_A(2,2) \big)  & = & 
     \mu \big( \cS_A(\half I; 2,2) \big) +  \mu\big( \cS_A(\half I; 2,2) \big) \, 
       m_1(0,\half)  \, \nu_2\big( SU(2) \big)  \nn \\
       & = & t \cdot 0 +  (1 \mm t) \cdot 0 \cdot 1 \cdot 1 = 0
  \qquad \qquad  \qed
\ee
independent of $t \in (0,1])$.
We can drop the requirement that $\gamma_A$ has rank 2 by 
observing that   extension to  all $  \gamma_{AB} $ of rank 2
 requires only that one replaces $(0,\half)$ on the right side of \eqref{gen22}
 by $[0,\half)$.  Thus, we can conclude that
 \begin{cor}
    \label{cor22}
If a state $\gamma_{AB}$ on ${\bf C}_2 \ot {\bf C}_2$ has rank 2,
then $\gamma_{AB}$ is almost surely entangled.
 \end{cor}

\subsection{Two-dimensional subspaces of ${\bf C}_{d}$.}   \label{sect:2tod}

We can  
 use the isomorphism between ${\bf C}_2$ and any Hilbert space
of dimension 2 to replace either $\cH_A$   or $\cH_B$  by a two 
dimensional subspace of ${\bf C}_d$.  
However, for later use, we now want to extend our qubit results to the 
somewhat more general situation of the set of all CPT maps 
$\Phi: {\bf C}_2 \mapsto {\bf C}_{d_B}$
whose range has the form $\cB \big(  \hbox{span}  \, \{  | v_1 \ket,  | v_2 \ket \} \big)$
with $| v_1 \ket,  |v_2 \ket \in  {\bf C}_{d_B}$.   Here, we do not fix the range,
but consider all CPT maps whose range corresponds to some two-dimensional
subspace of  ${\bf C}_{d_B}$.

  Observe that
 in the polar decomposition  $\Upsilon_{V^\dag} \circ \Phi \circ \Upsilon_U$  
 leading to the canonical form \eqref{KRrep1} we need only replace $V$ by an
 isometry $V: {\bf C}_2 \mapsto  {\cal H}_B$.
 Then in \eqref{cpt2} and \eqref{eb2}, the first use of $SU(2)$ in each subset
 must be replaced by $\cV_d $ which is defined as the subset of $d \times 2$
matrices satisfying $V^\dag V = I_2$.   By Theorem~A.2 of \cite{Arv},
$\cV_d$ can be given the structure of a real analytic manifold with a probaility
measure  $v_d$ (which is unique if it is required to be left-invariant under $SU(d)$).
Although  $\cV $ is not a group, we can define equivalence classes
as before with$  (V, U)  \simeq (V^\prime, U^\prime) $ if there is a $R_t \in {\cal R}_t $
 such that  $V^\prime = V R_t $ and $U^\prime = U R_t $.
  Then the previous arguments go through
with  $SU(2) \times SU(2)$ replaced by $\cV_d \times SU(2)$ in Section~\ref{sect:qq}
and the corresponding use of $\nu_2$ in Section~\ref{sect:meas2} by $v_d$.

\section{General maps}  \label{sect3}

\subsection{CPT maps with $d_A > 2$.}

We now assume $d_A \geq d_B \geq  2$ and
extend these results to bipartite states on ${\bf C}_{d_A} \ot {\bf C}_{d_B}$
with $\gamma_A = \tfrac{1}{d_A} I_A $.
We begin by considering a CPT  map $\Phi: M_{d_A} \mapsto M_{d_B} $ with Choi-rank
$d_A$.   By Theorem~5C of \cite{HSR}, which is equivalent to Corollary~\ref{cor:EBform},  
$\Phi$  can always be written in the form
\be
     \Phi(\rho) = \sum_k  \proj{g_k} \bra \psi_k,  \rho \, \psi_k \ket
\ee
where $\{ g_k \}$ is an orthonormal basis for ${\bf C}_{d_A} $, but the   states
$ \psi_k \in {\bf C}_{d_B}$ need {\em not} be orthogonal or even linearly independent.    
In the basis  $g_k$, the Choi matrix for $\Phi$ has the form
\be   \label{CJEB}
     \gamma_{AB} =   \tfrac{1}{d_A}  \sum_k \proj{g_k} \ot  \proj{\psi_k} 
\ee 
which implies that $\gamma_{AB}$  is block diagonal with each block a
$d_B \times d_B$ rank one projection.    Let us first assume that
 $\psi_1$ and $\psi_2$ are linearly independent.

Now let $P_k \equiv \proj{\psi_k}$ and write \eqref{CJEB} explicitly in block form,
as
\be
    \gamma_{AB} =  \tfrac{1}{d_A}  \pmx  P_1  & 0 & 0 & 0 & \ldots & 0 \\
           0 & P_2 & 0 & 0 & \ldots & 0  \\
           0 & 0 & P_3 & 0 & \ldots & 0   \\
           \vdots & & & \ddots & & \vdots \\
           0 & 0 & \ldots & 0 & \ldots & P_{d_A} \emx 
\ee
and consider a density matrix of the form %
\be   \label{Qform} 
     \tfrac{1}{d_A}  \pmx    Q & 0 & 0 & \ldots & 0 \\
            0 & P_3 & 0 & \ldots & 0   \\
           \vdots  & & \ddots & & \vdots \\
          0 & \ldots & 0 & \ldots & P_{d_A} \emx 
\ee
where $Q  \in M_2 \ot M_{d_B}$  is a positive semi-definite  $2 d_B \times 2 d_B$
matrix of rank 2 satisfying  $\trp_B  \, Q  =  I_2$. 
Now a density matrix of the form \eqref{Qform} is separable if and only if
$\half Q $ is separable.  However, $\half Q$ is a density matrix of the
form considered in Section~\ref{sect:2tod}.

Let   $  {\cal Y}_{d_A}(\{ g_k \} , \{ \psi_k \} ) $ denote the subset of 
$\cP_A(\tfrac{1}{d_A} I_{A}; d_A, d_A, )$ consisting of density matrices
of  the form  \eqref{Qform}
or, equivalently, 
\be
    {\cal Y}_{d_A}( \{ g_k \} , \{ \psi_k \} )  = \Big\{  
          Q \oplus \sum_{k = 3}^{d_A}    \proj{g_k} \ot  \proj{\psi_k} : Q \in  {\cal X}_{AB},
          \trp_B \, Q = I_2  \big\}
      \ee
      where $\oplus$ denotes the direct sum and
      \be
      {\cal X}_{AB} = \cB\big( \hbox{span} \{ |g_1 \ket, |g_2 \ket \}  \ot
           \cB\big( \hbox{span} \{  |\psi_1 \ket,  |\psi_2 \ket \}  \big).
      \ee
         The set of projections   $\proj{\psi_k} \in M_{d_B} $ is isomorphic to
     $S_{2d_B-1}$,   the $\ell_2$ unit sphere in ${\bf R}_{2d_B}$. 
        For a given $|g_1 \ket, |g_2 \ket $, the set  $ {\cal X}_{AB} $ depends only on
     span $ \{  |\psi_1 \ket,  |\psi_2 \ket \}$ and not the choice of individual vectors.  
    Therefore,   we can identify  each  point in 
      \be  \label{dCPTparam}
           \Omega_{d_A}  &  \equiv  &  \ovb{\Delta}_2 \times  \cV_{d_B}  \times SU(2) \times
             SU(d_A)/SU(2) \times  \underset{d_A - 2}
              {\underbrace{ S_{2d_B-1}  \times  \ldots  \times  S_{2d_B-1}  }} \nn \\
       & = &         \ovb{\Delta}_2 \times  \cV_{d_B}  \times SU(d_A) \times
              \underset{d_A - 2}
              {\underbrace{ S_{2d_B-1}  \times  \ldots  \times  S_{2d_B-1}  }} 
     \ee
     with a density matrix $\gamma_{AB} $  in   
   $         {\cal Y}_{d_A} \equiv \bigcup_{  \{ g_k \} , \{ \psi_k \} }
       {\cal Y}_{d_A}(\{ g_k \} , \{ \psi_k \} ) $, the set of all density matrices of
       the form \eqref{Qform}.
  (Note that $ S_{2d_B-1} $ occurs  $d_A - 2$
     times in \eqref{dCPTparam} corresponding to the choices 
     of $\psi_k$ for $k = 3 , 4 \ldots n$.   The set $ {\cal Y}_{d_A}(\{ g_k \} , \{ \psi_k \} ) $
     depends only on  span $ \{  |\psi_1 \ket,  |\psi_2 \ket \} = $
     range   $V$ with $V \in \cV_{d_B} $, with non-orthogonal vectors
    $|\psi_1 \ket,  |\psi_2 \ket $ associated with non-unital qubit channels
    via isomorphism.)
     
     Let $m_2$ and $v_d$ be measures as in Sections~\ref{sect:meas2}
     and \ref{sect:2tod}, let $\nu_d$ be normalized Haar measure
     on $SU(d)$ and let $n_{2d_B-1}$ be a  probability measure on $ S_{2d_B-1} $.
     We define a     normalized measure $\wtd{\mu}$ on  $\Omega_{d_A} $
    by the product measure  
     \be
         \wtd{\mu} =   m_2 \times v_{d_B} \times \nu_2 \times \nu_{d_A/2} \times 
            \underset{d_A - 2}
              {\underbrace{ n_{2d_B-1}  \times  \ldots  \times  n_{2d_B-1}  }}
     \ee   
     To obtain a measure on $  {\cal Y}_{d_A}$ we proceed as in
     Section~\ref{sect:meas2}.  Let  $G: \Omega_{d_A} \mapsto   {\cal Y}_{d_A}$ 
    be  the map that sends an element 
      $\big( (u,v), V, U,   |\psi_3\ket, \dots, |\psi_{d_A}\ket   \big)$
       to the corresponding density matrix in  $  {\cal Y}_{d_A}$ and define
     \be
         \mu(X)  = \wtd{\mu}\big( G^{-1}(X) \big)
     \ee
 whenever $X \subset  {\cal Y}_{d_A}$ for which $ G^{-1}(X)$ is measurable.
     
     As explained above, Corollary~\ref{cor:EBform} implies that 
      $\cS_A(\tfrac{1}{d_A} I_{A}; d_A, d_A ) \subset  {\cal Y}_{d_A}$.   
      Then, proceeding as in  \eqref{02}, one finds
\be   \label{d0}
    \mu\big( \cS_A(\tfrac{1}{d_A} I_{A}; d_A, d_A ) \big) = 
       0 \cdot 1 \cdot 1  \cdot 1^{d_A-2} = 0.
\ee     
   Moreover, for  any reasonable extension of $\mu$
from $ {\cal Y}_{d_A} $ to all  of $\cP_A(\tfrac{1}{d_A} I_{A},;d_A, d_A )$, the
EB subset will still have measure zero.   In particular, one could simply let
 \be  \label{finmeas}
    \omega(X) = \begin{cases}     \mu(X) & \quad \hbox{if} \quad X \subset  {\cal Y}_{d_A} \\
                0 & \quad \hbox{if} \quad X \subset  \cP_A(\tfrac{1}{d_A} I_{d}; d_A, d_A )
                  \backslash   {\cal Y}_{d_A}   \end{cases}
\ee
and note that $\omega$ is absolutely continuous with respect to any other extension
of $\mu$.

Thus, we have reduced the general case to that of $d_A = 2$
and conclude that
\begin{thm}   \label{lemmdd}
Let $\gamma_{AB}$ be a state on ${\bf C}_{d_A} \ot {\bf C}_{d_B}$ which has rank 
$d_A \geq d_B \geq 2$ and for which
$\gamma_A =  \tfrac{1}{d_A} I_{A}$.
  Then $\gamma_{AB}$ is almost surely entangled.
\end{thm}

{\bf Remark:}  The assumption that 
 $\psi_1$ and $\psi_2$  are linearly independent can be
 dropped because that case corresponds to $u = v = \tfrac{\pi}{2}$ in \eqref{KRrep1} 
 and is included implicitly in our analysis.    The set of channels for which all $\psi_j $
 are identical also has measure zero, except for the excluded situation $d_B = 1$,
 for which all states are separable. 
  

\subsection{Reduction of the general case to CPT}   \label{sect:reduct}

As observed earlier, when rank $\rho = d_A$
\be 
       \cP_A(\rho; d_A,d_A)   = 
            \big\{   (\sqrt{ d_A \, \rho} \ot I_B ) \gamma_{AB} (\sqrt{ d_A \, \rho} \ot I_B ) :
          \gamma_{AB} \in \cP_A(\tfrac{1}{d_A} I_A; d_A,d_A) \}
        \ee
        is isomorphic to $\cP_A(\tfrac{1}{d_A} I_A; d_A,d_A) $.        
But parameterizing the set of density matrices of rank $d_A$ is a bit more subtle
than for $d_A = 2$ because of the need to consider degenerate eigenvalues, for 
situations beyond $\tfrac{1}{d_A} I$.     However, this only affects a set of measure
zero and can be dealt with as in the preceding sections.
To describe the set of density matrix of rank $d_A$ consider the set of vectors
\be    \label{Z}
  Z =  \Big\{  {\bf z} = (\zeta_1, \zeta_2, \ldots \zeta_{d_A} ) : 0 < \zeta_1 \leq  \zeta_2 
      \leq  \ldots  \leq  \zeta_{d_A}, \sum_k \zeta_k = 1  \Big\}
\ee
in the positive facet of the $\ell_1$ unit ball of ${\bf R}_{d_A} $. 
We can associate each ${\bf z}  \in Z$ with a diagonal matrix $\Lambda_{\bf z}$ so that
that the map $h: ({\bf z}, U) \mapsto  U \Lambda_{\bf z} U^\dag$ takes $Z \times SU(d_A)$
onto $\cD_A(d_A)$, the set of density matrices in $M_{d_A}$ with full rank $d_A$.
Since we can identify $Z$ with a subset of ${\bf R}_{d_A -1}$, we
put normalized Lebesque measure $m_{d_A-1}$ on  $Z $, and let
\be   \label{eta}
    \eta_{d_A}(X) =  (m_{d_A-1} \times \nu_{d_A})( h^{-1}(X) )
\ee
whenever   $X \subset    \cD_A(d_A)$ and $h^{-1}(X)$ is measurable.
Then it follows from \eqref{d0} that for  any extension  $\omega$ of $\mu$,
the product measure $\omega \times \eta_{d_A}$ gives a measure
on $\cP_A(d_A,d_A)$ for which the separable states $\cS_A(d_A,d_A)$ have measure
$0 \cdot 1 = 0$.   Thus, we have proved 
\begin{thm}  \label{thmdd}
If a state $\gamma_{AB}$ on ${\bf C}_{d_A} \ot {\bf C}_{d_B}$  has rank $d_A \geq d_B \geq 2$
and  rank$(\gamma_A) = d_A$ then $\gamma_{AB}$ is almost surely entangled.
\end{thm}

\subsection{Further results}


Theorem~\ref{thm:r<s} states that if the rank of $\gamma_A$ is $d_A$ and
the rank of $\gamma_{AB} $ is strictly smaller than $d_A$, then
$\gamma_{AB}$ is entangled.   Thus  $r < d_A$ implies that $ \cP_A(\rho;r, d_A) $
consists entirely of entangled states.   If we combine this with our
results for $r = s = d_A$ we obtain several additional theorems, which we
state for completeness.

\begin{thm}   \label{thm:findA}
Assume $d_A \geq d_B \geq 2$.  If a state $\gamma_{AB}$ on $M_{d_A} \ot M_{d_B}$ has
rank $\gamma_{AB}  \leq d_A  = $ rank $ \gamma_A$,
 then $\gamma_{AB}$ is almost surely entangled.  
\end{thm} 

By using the isomorphism between  ${\bf C_d}$ and any Hilbert space of dimension $d$
we can restate this by letting $\cH_A = $ range $\gamma_A$ and 
$\cH_B = $ range $\gamma_B$ and considering $\gamma_{AB}$ as a state on
$\cB(\cH_A) \ot \cB(\cH_A)$.
\begin{thm}   \label{thm:finrA}
If a state $\gamma_{AB}$ on $M_{d_A} \ot M_{d_B}$ has
rank $\gamma_{AB}  \leq $ rank $ \gamma_A$ and 
rank $ \gamma_A  \geq $ rank $ \gamma_B \geq 2 $, then $\gamma_{AB}$
is almost surely entangled.  
\end{thm} 
 
We also find that we can eliminate the need to consider the rank of $\gamma_A$.
\begin{thm}  \label{thm:r_d_A}
Assume $d_A \geq d_B \geq 2$.  
If a state $\gamma_{AB}$ on $M_{d_A} \ot M_{d_B}$ has
rank $\gamma_{AB}  \leq d_A $, then $\gamma_{AB}$
is almost surely entangled.  
\end{thm} 
\pf  Let $\ovb{Z}$ denote the closure of  \eqref{Z}.   Since this simply replaces the 
strict inequality $0 < \zeta_1$ by $0 \leq \zeta_1$, the set 
 $\ovb{Z} \times SU(d)$ includes all density matrices in $M_{d_A}$ so that
\be
    \ovb{Z} \backslash Z \times SU(2) =  h^{-1}\big( \{ \rho \in \cP_A : ~  \hbox{rank} ~ \rho < d_A \} \big)
\ee
  Now extend the measure $\eta$ in \eqref{eta} to all of  $ \cD_A $.
The set of all separable states $\gamma_{AB} $ with rank $\gamma_{AB}  \leq d_A $
is   $\cS_A(d_A) \equiv  \bigcup_{s \leq  d_A} \cS_{d_A}(d_A,s) $.
The subset of separable states with rank  $\gamma_A < d_A $ satisfies
\be
 \bigcup_{s < d_A} \cS_{d_A}(d_A,s) 
  \subset \{ \rho \in \cP_A : ~  \hbox{rank} ~ \rho < d_A \} 
  \ee
But 
\be
   \eta_{d_A} \big(  \{ \rho \in \cP_A : ~  \hbox{rank} ~ \rho < d_A \} \big) = 
     m_{d_A-1} ( \ovb{Z} \backslash Z) ~  \nu_{d_A}(SU(2)) = 0 \cdot 1.
\ee
Thus
\be
   ( \omega_{d_A} \times \eta_{d_A}) \big(\cS(d_A) \big) & = &
       ( \omega_{d_A} \times \eta_{d_A})  \big( \cS_{d_A}(d_A,d_A) \big) +
           ( \omega_{d_A} \times \eta_{d_A})  \Big( \bigcup_{s < d_A} \cS_{d_A}(d_A,s)  \Big) \nn  \\
           & \leq &  0 \cdot 1 + 1 \cdot 0 = 0   \qquad \qquad \qed
\ee 

\section{Final comments}  \label{sect:fin}

\subsection{Remarks on measure}

If we apply the argument used in to prove Theorem~\ref{thm:r_d_A}
to the subset of states with $\gamma_A = \tfrac{1}{d_A} I_{d_A}$
or equivalently, combine  Theorems~\ref{lemmdd} and \ref{thm:r<s} we obtain the following
result which we state in terms of channels.
\begin{cor}   \label{cor}
Assume $d_A \geq d_B \geq 2$.  Then the set of CPT maps 
$\Phi : {\bf C}_{d_A} \mapsto {\bf C}_{d_B} $ whose Choi matrix
has rank $r \leq d_A$ is almost surely entanglement breaking.
\end{cor}
 
  As shown in Appendix~B, the closure of the set of extreme points of 
 CPT maps  $\Phi : {\bf C}_{d_A} \mapsto {\bf C}_{d_B} $ is precisely
the set of channels whose Choi matrix has rank $\leq d_A$.   Because
the extreme points of a convex set lie on the boundary, their closure
always has measure zero.  Thus,  Corollay 10
is  a special case of a well known,  more general fact from  convex geometry.  
An alternative, and somewhat simpler,  approach to proving 
Theorem~5, would be to use this fact together with Theorem~\ref{thm:ext}.
However, we feel that it is useful to see the specific paramaterizations
which lead to our results.   In our approach, one sees that everything
really follows from the basic paramaterization of extreme points for
qubit channels, and the fact 
that (up to sets of measure zero) the relevant sets of bipartite states
can be parameterized as direct products on which we can put
product measures.

One could extend Corollary~\ref{cor}  to the set of CP maps for which 
$\wh{\Phi}(I_B) = \rho$
with $\rho \in \cD_A(r)$ fixed, again using the fact that the closure of the
set of extreme points has measure zero.  Then we can conclude that the subset of
separable states $ \cup_{s \leq r} \cS(\rho;r,s)$ has measure zero with
respect to a measure on $ \cup_{s \leq r} \cP(\rho;r,s)$.     However, we can not go
directly from this observation to Theorem~\ref{thm:r_d_A} by taking
the $\bigcup_{\rho \in \cD_A} $ because the set $\cD_A$ is uncountable.
One would still need the argument in Section~\ref{sect:reduct}.
What this observation about extreme points does tell us is that our
results are not sensitive to the choice of measure.     The fundamental
issue is that the bipartite states can be parameterized as a smooth manifold
on which the separable ones correspond to a space of smaller dimensions.

There is one unsatisfying aspect   of using the the inverse image to define
a measure, as in \eqref{mudef}; namely, that it does not reflect the fact that different
unitaries give the same map on some lines in $\ovb{\Delta}$.   An alternative
would be to first define separate measures on  the different regions in \eqref{cpt2},
e.g.,  on $     \{ (u,u) \}_{u \in (0,\tfrac{\pi}{2}]} \times  (SU(2) \times SU(2))/{\cal R}_z  $,
use the product measure $m_1 \times \wtd{\nu}_z$ where $m_1$ is 
normalized Lebesque measure on $(0, \tfrac{\pi}{2)}$ and $\wtd{\nu}_z$   
is Haar measure on the group $ (SU(2) \times SU(2))/{\cal R}_z  $.   One could
then combine the measures on the four subsets in \eqref{cpt2} as in 
\eqref{m:gen2} using, say, weights  $1 - t_x - t_y - t_z,  t_z, t_x, t_y   $ with 
$t_m \geq 0$ and $\sum_{m=1}^3 t_m \leq 1$.   However, given that each of the
line segments with $u = \tfrac{\pi}{2}$, $u = v$, and $ v = \tfrac{\pi}{2}$ have
measure zero in $\ovb{\Delta}$, the most natural choice weight would be $t_m = 0$,
equivalent to simply omitting the corresponding channels (or states).

In fact, all regions  which a quotient space is needed, as in
Sections~\ref{sect:meas2}  and \ref{sect:reduct}
 have measure zero in our inverse image 
approach.    Intuitively, one would like to simply observe that we can identify
$\cD_A(\half I; 2,2)$ with a subset of $[0,  \tfrac{\pi}{2}] \times [0,  \tfrac{\pi}{2}]$
that satisfies  $ \Delta \subset \cD_A(\half I; 2,2) \subset  \ovb{\Delta} $ and then  
observe that since
\be
    \mu(\Delta) \leq  \mu\big( \cD_A(\half I; 2,2) \big) \leq  \mu(\ovb{\Delta} ).
\ee
and  $\mu(\Delta)  = \mu(\ovb{\Delta}) = 1$,  
one must have $ \mu\big( \cD_A(\half I; 2,2) = 1$.   But to use this approach,
one must  establish that  $\cD_A(\half I; 2,2)$ can be identified with a measurable
subset of $\ovb{\Delta}$.

\subsection{Optimality}
It is natural to ask if the results in Theorems~\ref{thm:findA} and \ref{thm:finrA} are optimal.
For $d_A > 2$, it is clear that the results which follow from
those of Walgate and Scott \cite{WS} are better.   Thus,
the question becomes whether or not  rank $\gamma_{AB} \leq (d_A - 1)(d_B - 1)$
is optimal.   This does not follow from the subspace theorems in \cite{WS} because when
rank $\gamma_{AB} = 2 > (d_A - 1)(d_B - 1)$  the product
states can form a set of measure zero in a subspace of $\cH_A \ot \cH_B$.
However, we know that the separable ball in $\cB(\cH_A \ot \cH_B)$ has strictly 
positive measure \cite{AS,GB,ZHSL} so that the optimal rank must be strictly smaller than
$d_A d_B$.

In the case of qubits, we know that Theorem~\ref{thm22} is stronger than the
results implied by Walgate and Scott \cite{WS}, and that when rank $\gamma_{AB} = 4$,
the separable states have strictly positive measure.    If we restrict attention to those
states $\gamma_{AB} $ with rank 3 and $\gamma_A = \gamma_B = \half I$ or, equivalently,
the unital CPT maps with Choi-rank 3, we can use the familiar picture of a
tetrahedron \cite{HH,EBQ,RSW}.   The rank  3 states correspond to the faces, and the
subset of separable states on each face to the smaller triangle whose vertices are
midpoints of the edges as shown in Figure~\ref{fig2}.
   Thus, the unital CPT maps with Choi-rank 3 have measure 
$0.25$  with respect to all the unital CPT maps on qubits.   However, it is open whether
or not this holds when the restriction to unital maps is removed.  Thus, the question
of whether $\cP_A(\half I; 3, 2)$ has measure zero or positive measure seems to be
open. 

 \begin{figure}[h]   \label{fig2}
 
 \begin{center}
  \includegraphics*[width=5cm,height=5cm,keepaspectratio=true]{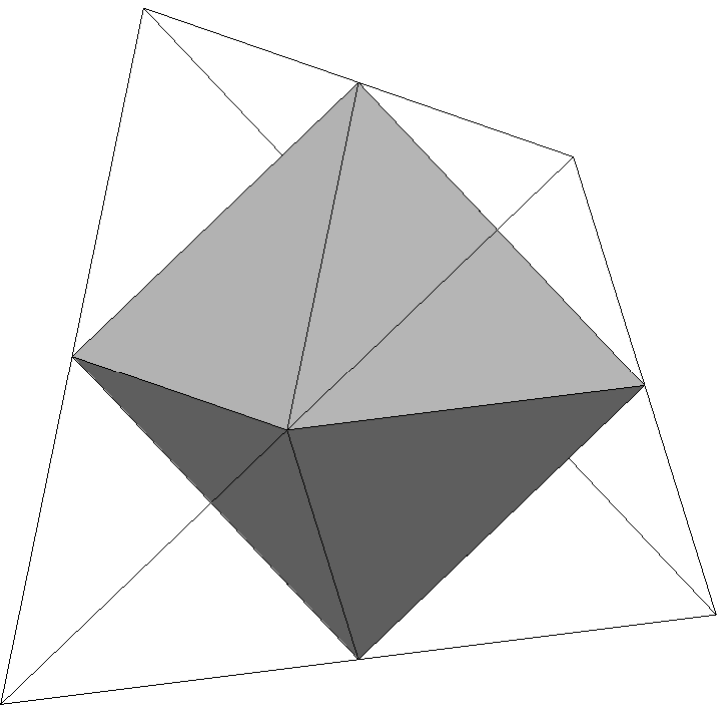}   
    \qquad 
     \includegraphics*[width=5cm,height=5cm,keepaspectratio=true]{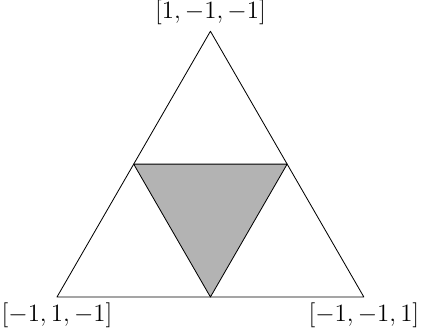}   
  \end{center}
  
 \caption{The left figure shows the tetrahedron of unital qubit channels with the
 octahedron the EB subset.   The right figure shows one of the faces of the tetrahedron, 
 corresponding to channels with Choi-rank 3, 
 with the shaded region the subset of EB channels}
 
 \end{figure}

\bigskip

{\bf Acknowledgment:}   It is a pleasure to thank Jonathon Walgate for bringing
\cite{WS} to our attention, to Michael Wolf for providing reference \cite{HSTT}
and to Michael Nathanson for assistance with figures.

\bigskip


 \appendix

 \section{Some separability  theorems}   \label{sect:A}
 
For completeness,  we now state and sketch proofs of some results that
are well known and/or proved in \cite{HSR}.    The first result appeared 
as \cite[Theorem 1]{HSTT} in a slightly stronger form.
 \begin{thm}   \label{thm:r<s}
 If rank $\gamma_{AB} <   d_A =  ~ \hbox{rank} ~\gamma_A$, then $\gamma_{AB} $ is
 not separable.  
 \end{thm}
 \pf    First observe that $\gamma_{AB} $ is separable if and only if
 \be
     \wtd{\gamma}_{AB} \equiv  \tfrac{1}{d_A} (I_A \ot \gamma_A)^{-1/2} 
          \gamma_{AB}   (I_A \ot \gamma_A)^{-1/2}  
 \ee
is separable.   But $\wtd{\gamma}_A =  \tfrac{1}{d_A} I_A $.   Now both
the reduction and majorization   criteria  \cite{B,H4rev} for separability of a state $\rho_{AB} $
imply that the largest eigenvalue must satisfy
  $ \norm{ \rho _{AB}}_\infty  \leq   \norm{ \rho_{B}}_\infty $.
But  rank $\wtd{\gamma}_{AB} = $ rank $   \gamma_{AB} < d_A$ implies that
$\wtd{\gamma}_{AB}$ has at least one eigenvalue $ >  \tfrac{1}{d_A} $.  Thus 
$    \norm{ \wtd{\gamma} _{AB}}_\infty >  \tfrac{1}{d_A}  =    \norm{ \wtd{\gamma} _{B}}_\infty $,
and it follows that both $ \wtd{\gamma}_{AB}$ and $\gamma_{AB} $ are entangled.  \qed

When rank $\gamma_A < d_A$, one can regard the underlying Hilbert space
as ${\cal H}_A$ to be range~$\gamma_A = (\ker \gamma_A)^\perp$.
One then obtains 
\begin{cor} \label{cor:r<s}
 If rank $\gamma_{AB} <  $ rank $ \gamma_A$, then $\gamma_{AB} $ is
 not separable.  
\end{cor} 

The following Lemma goes back at least to \cite{HLVC} and a
simpler proof was given in \cite{HSR}.   To emphasize 
that one need not assume  $d_A = d_B$ (and because
of typos in \cite{HSR}) we include a full proof here.
\begin{lemma}  \label{lemm:dsep}
Let $\rho_{AB} $ be a density matrix on
${\cal H}_A \ot {\cal H}_B$.  If $\rho_{AB} $ is separable,
$\rho_{AB} $ has rank $d$, and
$\rho_A $ has rank $d$, then $\rho_{AB} $ can
be written as a convex combination of products of pure
states using at most $d$ products.
\end{lemma}    \label{lemm:CQ}
\pf Since $\rho_{AB} $ is separable it can be written in the form
\be   \label{eq:minprod}
   \rho_{AB}= \sum_{i=1}^k \lambda_i \, \proj{a_i} \otimes \proj{b_i}.
\ee
with $\norm{ a_i} = \norm{b_i} = 1$.
Assume that $k > d$ and that $\rho_{AB} $ can not be written in
the form (\ref{eq:minprod}) using less than $k$ products.
Since $\rho_A$ has exactly rank $d$, there is no loss of
generality in assuming that the vectors above
have been chosen so that $|a_1 \ket, |a_2 \ket, \ldots |a_d \ket$
are linearly independent.   Moreover,
since $\rho_{AB} $ has rank $d<k$,
the first $d+1$ vectors $|a_i \ket \otimes |b_i\ket $
must be linearly dependent so that one can find $\alpha_j$  such that
\be
\sum_{j=1}^{d+1} \alpha_j \, |a_j \ket \ot  |b_j\ket  = 0.
\ee
Now let $\{ |e_k \ket \}$ be an orthonormal basis for ${\cal H}_B$.
Then
\be   \label{eq:ldA}
\sum_{j=1}^{d+1} \alpha_j \bra e_k, b_j \ket \, |a_j \ket   = 0 ~~~\forall ~k.
\ee
Since the first $d$ vectors $|a_j \ket $ are linearly
independent, the solution of   $\sum_j x_j |a_j \ket = 0$ 
is unique up to a multiplicative constant.  
Applying this to the coefficients in (\ref{eq:ldA})
one finds that there are numbers
$\nu_k$ such that $\alpha_j \bra e_k, |b_j \ket = \nu_k x_j$.
Let $|\nu \ket \equiv \sum_k \nu_k |e_k \ket$.
Then $\alpha_j |b_j \ket = x_j |\nu \ket$.  
Since multiplying  $x_j$ by $c$, changes $\nu_k \raw \tfrac{1}{c} \nu_k$,
one can assume that $x_j$  has been chosen so that 
$ \norm{\nu} = 1 =  \norm{b_j} $.   Then
$\alpha_j |b_j \ket = x_j e^{i \theta_j} | \nu \ket$, and
$ \alpha_j \neq 0$ implies $|b_j \ket = e^{i  \theta_j} | \nu \ket$.
Therefore, , one can rewrite (\ref{eq:minprod}) as
\be  \label{eq:alt.rho}
   \rho_{AB}= \sum_{j : \alpha_j = 0} \lambda_j \,
        \proj{a_j} \otimes \proj{b_j}  + \sum_{j : \alpha_j \neq 0}
    \lambda_j  \, \proj{a_j} \otimes \proj{\nu} .
\ee
Suppose that $t$ of the $\alpha_j$ are non-zero.
Since the vectors $\{ | a_j \ket : \alpha_j \neq 0 \}$ are
linearly dependent, the density matrix
$ \sum_{j : \alpha_j \neq 0}  \lambda_j  \, \proj{a_j}$
has rank strictly $< t$ and can be rewritten in the form
$\sum_{k = 1}^{s} \lambda_j^\prime \proj{a_j^\prime}$
using only $s < t$ vectors $|a_j^\prime\ket$ .  Substituting this in (\ref{eq:alt.rho})
gives $\rho_{AB} $ as linear combination of products using strictly
less than $k$ contradicting the assumption that (\ref{eq:minprod})
used the minimum number.   \qed

\begin{cor}  \label{cor:EBform}
 If $\gamma_{AB}$ is separable and $\gamma_A = \tfrac{1}{d_A} I_A$, then
 $\gamma_{AB}$ can be written in the form 
 \be   \label{eq:EBform}
    \gamma_{AB} = \sum_k \tfrac{1}{d_A} \proj{g_k} \ot \proj{\psi_k}
 \ee
 with $g_k$ an orthonormal basis for ${\bf C}_{d_A} $
\end{cor}
\pf  Since $\gamma_{AB}$ is separable it is a convex
combination of projections onto product states and can
be written in the form
\be  \label{EBcor2}
    \gamma_{AB} = \sum_k  \xi_k \proj{g_k \ot \psi_k}.
\ee
Since rank $\gamma_A$ is $d_A$ by assumption, it follows
from Lemma~\ref{lemm:CQ}  that we can assume $k = 1, 2 \ldots d_A$
(duplicating terms if $< d_A$ are needed).  But then, the assumption
\be
      \tfrac{1}{d_A} I_A = \gamma_A = \sum_k   \xi_k \proj{g_k} 
\ee
 holds if and only if $\xi_k =  \tfrac{1}{d_A}  ~~ \forall ~k$ and
the vectors $g_k$ are orthonormal.  \hskip1cm  \qed
 
 \section{Closure of the set of extreme points}  \label{app:ext}
 
  It is often useful to consider the set of all CPT maps with 
Choi rank $\leq d_A$.   In \cite{RSW} these were called   ``generalized extreme points''
and shown to be equivalent to the closure of the set of extreme points
for qubit maps.    This is true in general.\footnote{Arveson  \cite{Arv3} has pointed out that
Theorem~\ref{thm:ext} can also  be proved using results  in \cite{Arv}.}   We repeat
here an argument form \cite{Ropen}.
Let $\cE(d_A,d_B)$ denote the extreme points of the convex set of 
CPT maps from $M_{d_A}$ to $M_{d_B}$.  
\begin{thm}   \label{thm:ext}
The closure
$\ovb{\cE(d_A,d_B)}$ of the set of extreme points of CPT maps
$\Phi: M_{d_A}  \mapsto M_{d_B}$ is precisely the set of such maps
with Choi rank  at most $d_A$.   
  \end{thm}
\pf     Let $A_k$ be the Choi-Kraus operators for a
map $\Phi: M_{d_A}  \mapsto M_{d_B}$ with Choi rank $r \leq d_A$ which
is not extreme, and let $B_k$ be the Choi-Kraus operators for a true extreme point
with Choi-rank $d_A$.
When $r < d_A$ extend $A_k$ by letting $A_m = 0$ for $m = r \pp 1, r \pp 2, \ldots d_A$
and define $C_k(\e) = A_k + \e B_k$.     There is a number $\e_* $ such that 
the $d_A^2$ matrices  $C_j^\dag(\e) C_k(\e) $ are linear independent
for $0  <  \e < \e_* $.     To see this,  for each $C_j^\dag(\e) C_k(\e) $ ``stack'' 
the columns to give a  
vector of length $d_A^2$ and let  $M(\e)$ denote the  $d_A^2  \times d_A^2$
  matrix formed with these vectors as columns.   Then $\det M(\e)$ is a polynomial of
degree $d_A^4$, which has  at most  $d_A^4$ distinct roots.  Since the matrcies
$A_j^\dag A_k$ were assumed to be linearly dependent, one of these 
roots is $0$; it suffices to take  $\e_*$ the next largest root (or $+1$ if no 
roots are positive).  
Thus, the operators $C_j^\dag(\e) C_k(\e) $ are linearly independent  for $\e \in (0, \e_* )$.
The map  $\rho \mapsto \sum_k C_k(\e) \rho \, C_k^\dag(\e)$ is CP,
with  
\bee \sum_k C_k^\dag(\e)  \, C_k(\e) = (1 + \e^2)I + \e( A_k^\dag B_k + B_k^\dag A_k) \equiv S(\e).
\eee
For sufficiently small $\e$ the operator $S(\e)$ is positive semi-definite and
invertible, and the map $\Phi_{\e}(\rho) =  C_k(\e) S(\e)^{-1/2} \rho   S(\e)^{-1/2} C_k^\dag(\e)$ 
is a CPT map with Kraus operators $C_k(\e) S(\e)^{-1/2}  $.     Thus,
one can find $\e_c$ such that $\e \in (0,\e_c)$ implies that 
 $\Phi_{\e} \in \cE(d_A,d_B)$.   It then follows from
  $\ds{ \lim_{\e \raw 0+} \Phi_{\e} = \Phi}$ that
 $\Phi \in  \ovb{\cE(d_A,d_B)}$.     \qed

 \bigskip
 

\end{document}